\documentclass[fleqn,usenatbib]{mnras}

\usepackage[T1]{fontenc}

\DeclareRobustCommand{\VAN}[3]{#2}
\let\VANthebibliography\thebibliography
\def\thebibliography{\DeclareRobustCommand{\VAN}[3]{##3}\VANthebibliography}

\usepackage{graphicx}	
\usepackage{amsmath}	
\usepackage{amssymb}	
\usepackage{newtxtext,newtxmath}

\title[Detection of HNCN in the ISM]{Detection of the cyanomidyl radical (HNCN): a new interstellar species with the NCN backbone}

\author[V. M. Rivilla et al.]{
V. M. Rivilla,$^{1,2}$\thanks{E-mail: vrivilla@cab.inta-csic.es}
I. Jim\'enez-Serra,$^{1}$
J. Garc\'ia de la Concepci\'on,$^{1}$
J. Mart\'in-Pintado,$^{1}$
L. Colzi,$^{1}$
\newauthor 
L. F. Rodr\'iguez-Almeida,$^{1}$
B. Tercero,$^{3}$
F. Rico-Villas,$^{1}$
S. Zeng$,^{4}$
S. Mart\'in,$^{5,6}$
M. A. Requena-Torres,$^{7,8}$
\newauthor
P. de Vicente,$^{3}$
\\
$^{1}$Centro de Astrobiolog\'ia (CSIC-INTA), Ctra. de Ajalvir Km. 4, Torrej\'on de Ardoz, 28850 Madrid, Spain\\
$^{2}$INAF-Osservatorio Astrofisico di Arcetri, Largo Enrico Fermi 5, 50125, Florence, Italy\\
$^{3}$Observatorio Astronómico Nacional (OAN-IGN), Calle Alfonso XII, 3, 28014 Madrid, Spain\\
$^{4}$RIKEN, 2-1 Hirosawa, Wako, Saitama, 351-0198, Japan\\
$^{5}$European Southern Observatory, ALMA Department of Science, Alonso de Córdova 3107, Vitacura 763 0355, Santiago, Chile\\
$^{6}$Joint ALMA Observatory, Department of Science Operations, Alonso de Córdova 3107, Vitacura 763 0355, Santiago, Chile\\
$^{7}$University of Maryland, Department of Astronomy, College Park, ND 20742-2421, USA\\
$^{8}$Department of Physics, Astronomy and Geosciences, Towson University, Towson, MD 21252, USA\\
}

\date{Accepted 2021 June 17. Received 2021 June 12; in original form 2021 May 28.}

\pubyear{2015}

\begin{document}
\label{firstpage}
\pagerange{\pageref{firstpage}--\pageref{lastpage}}
\maketitle

\begin{abstract}
We report here the first detection in the interstellar medium of the cyanomidyl radical (HNCN). Using the Yebes 40m and the IRAM 30m telescopes, we have targeted the doublets of the $N$=2$-$1, 4$-$3, 5$-$4, 6$-$5, and 7$-$6 transitions of HNCN toward the molecular cloud G+0.693-0.027. 
We have detected three unblended lines of HNCN, these are the $N$=6$-$5 doublet and one line of the $N$=4$-$3 transition. Additionally we present one line of the $N$=5$-$4 transition partially blended with emission from other species.
The Local Thermodynamic Equilibrium best fit to the data gives a molecular abundance of (0.91$\pm$0.05)$\times$10$^{-10}$ with respect to H$_2$. 
The relatively low abundance of this species in G+0.693-0.027, and its high reactivity, suggest that HNCN is possibly produced by gas-phase chemistry. Our work shows that this highly reactive molecule is present in interstellar space, and thus it represents a plausible precursor of larger prebiotic molecules with the NCN backbone such as cyanamide (NH$_2$CN), carbodiimide (HNCNH) and formamidine (NH$_2$CHNH).
\end{abstract}




\section{Introduction} 
\label{sec:intro}

\begin{figure}
\hspace{-0.6cm}
\includegraphics[scale=0.425]{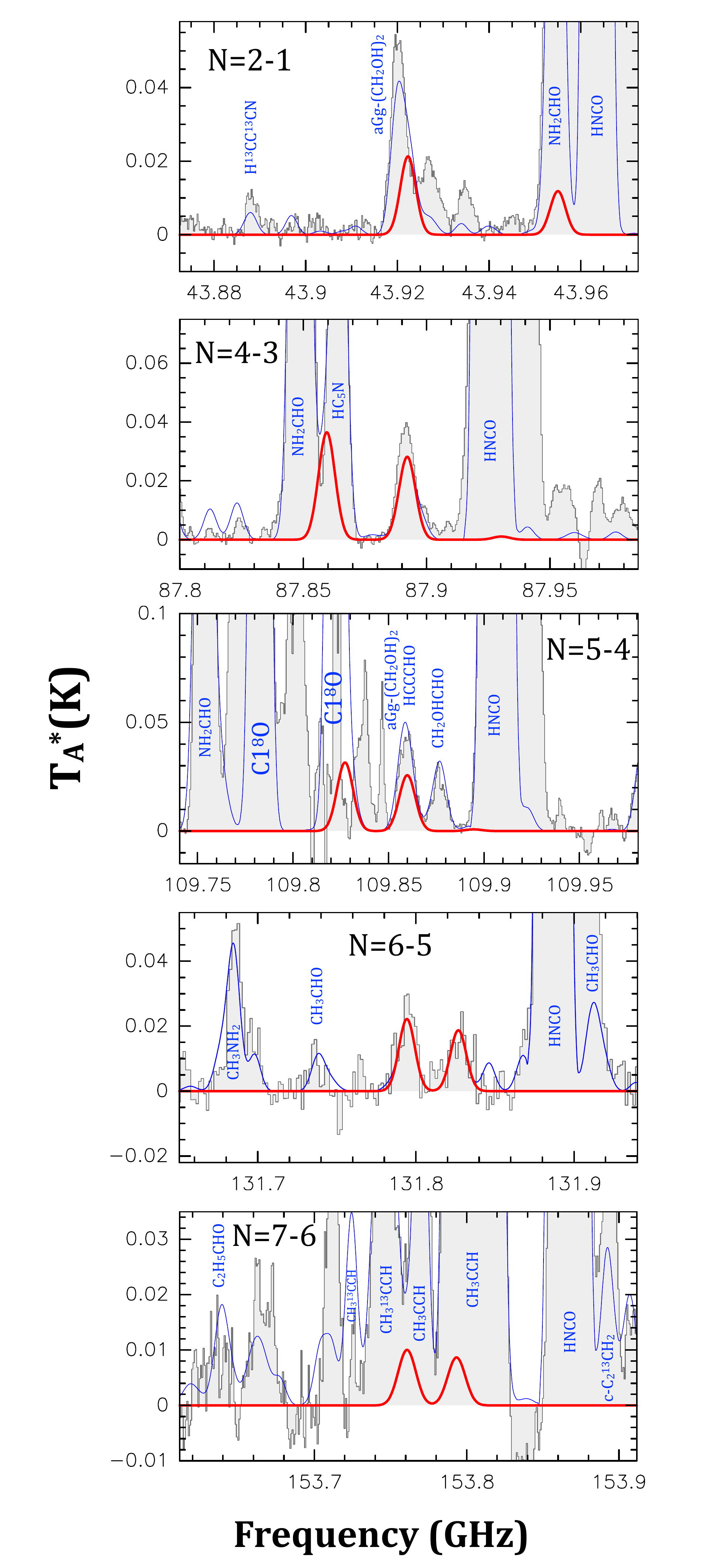}
\centering
\vspace{-3mm}
\caption{Transitions of HNCN detected in the observed spectra of G+0.693. The best LTE fit to the HNCN lines is shown in red, while the total contribution, including other molecular species (labelled) identified in the region is shown in blue.}
\label{fig-hncn}
\end{figure}

The study of the molecular complexity of the interstellar medium (ISM) provides us fundamental information on the chemical reservoir available for the formation of stars and planets. It also gives us hints about the chemical inventory that the primitive Earth may have inherited.
Molecular species with the nitrogen-carbon-nitrogen (NCN) backbone are especially relevant for prebiotic chemistry, since e.g. the heterocyclic rings of RNA and DNA nucleobases.
 
In the last decade, many prebiotic experiments have shown that simple species with the NCN backbone such as cyanamide (NH$_2$CN, \citealt{powner2009,fahrenbach2017,kaur2019}), its isomer carbodiimide (HNCNH; \citealt{williams1981,merz2014}),
and urea (NH$_2$CONH$_2$, \citealt{sanchez1966,becker2019}) are key molecular precursors for the synthesis of RNA and DNA ribonucleotides under primitive Earth conditions. Interestingly, all these three species have been detected in the ISM (\citealt{turner1975,macguire2012,belloche2019}). However, the simpler cyanomidyl radical (HNCN) has so far eluded detection in space (\citealt{yamamoto1994}). This radical can play an important role in the chemistry of molecules containing the NCN backbone, and thus its detection in the ISM would contribute to a better understanding of the potential formation of relevant prebiotic species through interstellar chemistry.
With this goal we searched for HNCN towards the G+0.693-0.027 molecular cloud (G+0.693, hereafter) located in the  Sgr B2 complex in the Galactic Center. This cloud is the ideal target to detect new molecular species since it shows a very rich chemistry in complex organic molecules (\citealt{requena-torres_largest_2008,zeng2018,rivilla2019b,rodriguez-almeida2021}), and, in particular, in nitrogen-bearing species, including cyanamide (NH$_2$CN, \citealt{zeng2018}), hydroxylamine (NH$_2$OH, \citealt{rivilla2020b}), urea (NH$_2$CONH$_2$, \citealt{jimenez-serra2020}), iminopropyne (HCCCHNH, \citealt{bizzocchi2020}), and ethanolamine (NH$_2$CH$_2$CH$_2$OH, \citealt{rivilla2021}). In this Letter, we present the first detection of HNCN in the ISM towards G+0.693 through the detection of several rotational transitions of its millimeter spectrum.

\vspace{-6mm}
\section{Observations} 
\label{sec:observations}

We have carried out a high-sensitivity spectral survey from 31 to 278 GHz towards the G+0.693 molecular cloud using the Yebes 40m telescope (Guadalajara, Spain) and the IRAM 30m telescope (Granada, Spain). The observations were centered at $\alpha$(J2000.0)=$\,$17$^{\rm h}$47$^{\rm m}$22$^{\rm s}$, $\delta$(J2000.0)=$\,-$28$^{\circ}$21$^{\prime}$27 $^{\prime\prime}$.

\begin{table}
\centering
\tabcolsep 2.5pt
\caption{List of observed transitions of HNCN. We provide the parameters from the CDMS catalog entry 041507: frequencies, quantum numbers, and upper state degeneracy (g$_{\rm u}$). We also show the values of the base 10 logarithm of the Einstein coefficients (log $A_{\rm ul}$), and the energy of the upper levels ($E_{\rm u}$). The last column indicates the molecular species that produces blending with some HNCN transitions.}
\begin{tabular}{l c c  c c l}
\hline
 Frequency & Transition    & log  $A_{\rm ul}$  & $g_{\rm u}$ & E$_{\rm u}$ &   Blending \\
 (GHz) & ($N_{\rm K,k,J}$)   &  (s$^{-1}$) & & (K) &   \\
\hline
43.9225164$^*$    & 2$_{0,2,5/2}-$1$_{0,1,3/2}$    & -5.688  &   6 &  3.2 &  aGg-(CH$_2$OH)$_2$   \\ 
43.9552012    & 2$_{0,2,3/2}-$1$_{0,1,1/2}$    & -5.7663 &   4 &  3.2 &  NH$_2$CHO  \\
87.8601563    & 4$_{0,4,9/2}-$3$_{0,3,7/2}$     & -4.739  &  10 &  10.5 &  HC$_5$N  \\
87.8927924$^*$    & 4$_{0,4,7/2}-$3$_{0,3,5/2}$   & -4.7543  &  8 &  10.5 & unblended  \\
109.8281277   & 5$_{0,5,11/2}-$4$_{0,4,9/2}$    & -4.4385 &   12 &  15.8 &  C$^{18}$O \\
109.8607271$^*$    & 5$_{0,5,9/2}-$4$_{0,4,7/2}$    & -4.4479 &   10 &  15.8 &   aGg-(CH$_2$OH)$_2$, HCCCHO \\
131.7953263$^*$    & 6$_{0,6,13/2}-$5$_{0,5,11/2}$     & -4.1943 &   14 &  22.1 & unblended  \\
131.8278808$^*$    & 6$_{0,6,11/2}-$5$_{0,5,9/2}$    & -4.2006 &   12 &  22.1 &  unblended \\
153.7615965   & 7$_{0,7,15/2}-$6$_{0,6,13/2}$    & -3.9886 &   16 &  29.5 &   CH$_3^{13}$CCH, CH$_3$CCH\\
153.7940979  & 7$_{0,7,13/2}-$6$_{0,6,11/2}$    & -3.9932 &   14 &  29.5 &  CH$_3$CCH \\
\hline 
\end{tabular}
\label{tab:transitions}
\vspace{-2mm}
{\\ $^*$ Transitions used to perform the fit with the SLIM-AUTOFIT tool of MADCUBA (see text).}
\end{table}

The Yebes 40m observations were carried out during 6 observing sessions in February 2020, as part of the project 20A008 (PI Jim\'enez-Serra). We used the Nanocosmos Q-band (7 mm) HEMT receiver that allows ultra broad-band observations in two linear polarizations (\citealt{tercero2021}). The receiver is connected to 16 fast Fourier transform spectrometers (FFTS) with a spectral coverage of 2.5 GHz and a spectral resolution of 38 kHz.
The total spectral coverage was from 31.075 GHz to 50.424 GHz. 
The final spectra were smoothed to a resolution of 251 kHz, corresponding to a velocity resolution of 1.7 km s$^{-1}$ at 45 GHz.
The position switching mode was used, with the off position located at (-885$^{\prime\prime}$,+290$^{\prime\prime}$) with respect to G+0.693, as in previous works (e.g. \citealt{rivilla2020b}). The telescope pointing and focus were checked every one or two hours through pseudo-continuum observations towards the red hypergiant star VX Sgr. The spectra were measured in units of antenna temperature, T$_{\rm A}^{*}$, and corrected for atmospheric absorption and spillover losses. 
The noise of the spectra depends on the frequency, spanning in the range 1$-$4  mK. 
The half power beam width (HPBW) of the telescope is $\sim$39$^{\prime\prime}$ at 44 GHz.

The IRAM 30m observations were performed in three observing runs during 2019 (April 10$-$16, August 13$-$19, and December 11$-$15), as part of the projects 172$-$18 (PI Mart\'in-Pintado), 018$-$19 and 133$-$19 (PI Rivilla). We used the broad-band Eight MIxer Receiver (EMIR) and the fast Fourier transform spectrometers in FTS200 mode, which provided a channel width of $\sim$200 kHz.
The covered spectral ranges were 71.76$-$116.72$\,$GHz, 124.77$-$175.5$\,$GHz, and 223.307$-$238.29 GHz.
The final spectra were smoothed to 609 kHz, i.e. a velocity resolution of 1.8 km s$^{-1}$ at 100 GHz.
The telescope pointing and focus were checked every 1.5 h towards bright sources. The spectra were measured in the T$_{\rm A}^{*}$ scale. 
The noise of the spectra depends on the frequency range. Approximate values are:  1.7$-$2.8 mK ($\sim$71-90 GHz), 1.5$-$9.8 mK ($\sim$90$-$115 GHz), 3.1$-$6.8 mK ($\sim$124$-$175 GHz), and 4.7$-$9.7 mK (223$-$238 GHz).   
The HPBW of the observations vary between 10.3$^{\prime\prime}$ and 21.6$^{\prime\prime}$.
We used the same off position for the position switching observations as in the Yebes 40m observations.

\vspace{-5mm}
\section{Analysis and Results} 
\label{sec:analysis}

The identification of the molecular lines of HNCN was performed using the SLIM (Spectral Line Identification and Modeling) tool within the MADCUBA package{\footnote{Madrid Data Cube Analysis on ImageJ is a software developed at the Center of Astrobiology (CAB) in Madrid; http://cab.inta-csic.es/madcuba/Portada.html.}} (version 21/12/2020; \citealt{martin2019}). We used the spectroscopic entry 041507 (July 2012) from the Cologne Database for Molecular Spectroscopy (CDMS, \citealt{endres2016}), which was obtained from the laboratory work by \citet{yamamoto1994}. The ab initio dipole moment components were calculated by H. S. P. M\"uller (2012, unpublished).
The lines available in this entry corresponds to the K structure of the a-type R branch transitions of a slightly asymmetric rotor (\citealt{yamamoto1994}), and considers fine structure splitting (denoted by the $J$ quantum number).
We have targeted the transitions $N$=2$-$1, 4$-$3, 5$-$4, 6$-$5 and 7$-$6, which splits each into three lines. Only the doublets with $\Delta$J$\neq$0 are bright enough to be detected (see Table \ref{tab:transitions}). 
Given the low excitation temperatures found in G+0.693 of $T_{\rm ex}\sim$5$-$20 K (\citealt{requena-torres_largest_2008,zeng2018,Rivilla2018,rivilla2019b}), these low energy transitions are the only ones expected to produce significant emission, with energies of their upper level in the range 3$-$30 K (Table \ref{tab:transitions}).

For the identification of HNCN, we compared the synthetic spectra generated by SLIM (red curve in Fig. \ref{fig-hncn}), under the assumption of Local Thermodynamic Equilibrium (LTE) conditions, with the observed spectra (black histogram and grey-shaded area; Fig. \ref{fig-hncn}). To evaluate if the transitions are blended with emission from other species, we have also considered the LTE model that predicts the total contribution of all the species that our group has identified so far towards G+0.693 ($>$120), including HNCN (blue curve in Fig. \ref{fig-hncn}).
Three of the transitions of HNCN, namely the doublet of the $N$=6$-$5 transition and one of the $N$=4$-$3 lines, are detected in the observed spectra without being blended with other species (Fig. \ref{fig-hncn}). The detection of both members of the $N$=6$-$5 rotational transition provides solid support for the unambiguous identification of HNCN in the observed spectra.
One of the $N$=5$-$4 doublets appears only partially blended with emission of ethylene glycol (agG-(CH$_2$OH)$_2$) and propynal (HCCCHO). The profile predicted considering the emission from all species matches well the observed spectrum (middle panel in Fig. \ref{fig-hncn}). The remaining transitions of HNCN are blended with bright emission from other species already identified in the source. We indicate the contaminant species in the last column of Table \ref{tab:transitions}.

We note that the $N$=5$-$4 transition centered at 108.82 GHz is blended with C$^{18}$O  $J$=1$-$0, which shows multiple velocity components along the line of sight toward this position in the Galactic Center (see Appendix A). The narrow C$^{18}$O $J$=1$-$0 feature seen at this frequency (linewidths of 5 km s$^{-1}$, i.e. a factor of $\sim$3$-$5 narrower than typically observed toward G+0.693) is an artefact produced by the subtraction of the spectrum measured toward the off position, which presents emission only from CO and its isotopologues (see Appendix A).

\begin{table}
\centering
\tabcolsep 1.5pt
\caption{Derived physical parameters of HNCN and other molecular species that might be chemically related.}
\begin{tabular}{ c  c c c c c c  }
\hline
 Molecule & $N$   &  $T_{\rm ex}$ & v$_{\rm LSR}$ & $FWHM$  & Abundance$^a$ & Ref.$^b$  \\
 & ($\times$10$^{13}$ cm$^{-2}$) & (K) & (km s$^{-1}$) & (km s$^{-1}$) & ($\times$10$^{-10}$) &   \\
\hline
HNCN  & 1.2$\pm$0.1  & 8.7$\pm$0.7  & 71$\pm$1  & 25 & 0.91$\pm$0.05 &  (1)  \\  
HNCNH &  $<$2.5 &  8 & 69 & 25 & $<$1.9 &  (1) \\
NH$_2$CN$^c$  & 27$\pm$2 & 6.8$\pm$0.2 & 67$\pm$1  & 24$\pm$1   & 23.0  &  (2)  \\
\hline 
\end{tabular}
\label{tab:parameters}
\vspace{0mm}
{\\ (a) We adopted $N_{\rm H_2}$=(1.4$\pm$0.3)$\times$10$^{23}$ cm$^{-2}$, from \citet{martin_tracing_2008}.
(b) References: (1) This work; (2) \citet{zeng2018}; 
(c) $N$, v$_{\rm LSR}$ and $FWHM$ of the para species; the abundance corresponds to the sum of the ortho and para species.
}
\end{table}

To derive the physical parameters of the HNCN emission we used the AUTOFIT tool of SLIM, which finds the best agreement between the observed spectra and the predicted LTE model (see details in \citealt{martin2019}). For the fit we considered not only the emission of HNCN, but also the predicted emission from all the species identified in this source.
We have used the HNCN transitions that appear unblended or only partially blended, and also the low-excitation $N$=2$-$1 transition at 43.9225164 GHz (upper panel in Fig. \ref{fig-hncn}), which, although blended, allows us to better constrain the excitation temperature. We fixed the linewidth (full width at half maximum, $FWHM$) to 25 km s$^{-1}$, which reproduces well the spectral profile of the unblended transition, and is similar to those derived for other N-bearing species (see e.g. \citealt{zeng2018}). We left free the other parameters: column density ($N$), excitation temperature ($T_{\rm ex}$) and velocity ($v_{\rm LSR}$). The best fit gives $N$=(1.1$\pm$0.1)$\times$10$^{13}$ cm$^{-2}$, $T_{\rm ex}$=8.7$\pm$0.7 K, and a velocity of $v_{\rm LSR}$=71.3$\pm$0.1 km s$^{-1}$. The derived $T_{\rm ex}$ and $v_{\rm LSR}$ are consistent to those observed for other species observed previously in G+0.693 (\citealt{requena-torres_largest_2008,zeng2018,rivilla2019b,rivilla2020b,jimenez-serra2020}). To derive the abundance of HNCN with respect to H$_2$, we have used the H$_2$ column density inferred from observations of C$^{18}$O (\citealt{martin_tracing_2008}), obtaining a value of (0.91$\pm$0.05)$\times$10$^{-10}$. The derived physical parameters of the HNCN emission are summarised in Table \ref{tab:parameters}.

We have also searched for other molecular species that are expected to be chemically related to HNCN. 
Cyanamide (NH$_2$CN) was reported by \citet{zeng2018}, with an abundance of 23$\times$10$^{-10}$ (Table \ref{tab:parameters}). We did not detect in our survey carbodiimide (HNCNH), an isomer of cyanamide. 
To derive an upper limit of the HNCNH abundance we used the brightest transition predicted by the LTE model at 8 K within our survey, which is the 6$_{1,6}-$7$_{0,7}$ transition at 223.791 GHz (E$_{\rm up}$=38.6 K).
We obtained an abundance of $<$1.9$\times$10$^{-10}$ (Table \ref{tab:parameters}), more than an order of magnitude less abundant than cyanamide.

\vspace{-5mm}

\section{Discussion}
\label{sec:discussion}

\subsection{Formation of HNCN in the ISM}
\label{sec:formation}

\begin{figure}
\includegraphics[scale=0.185]{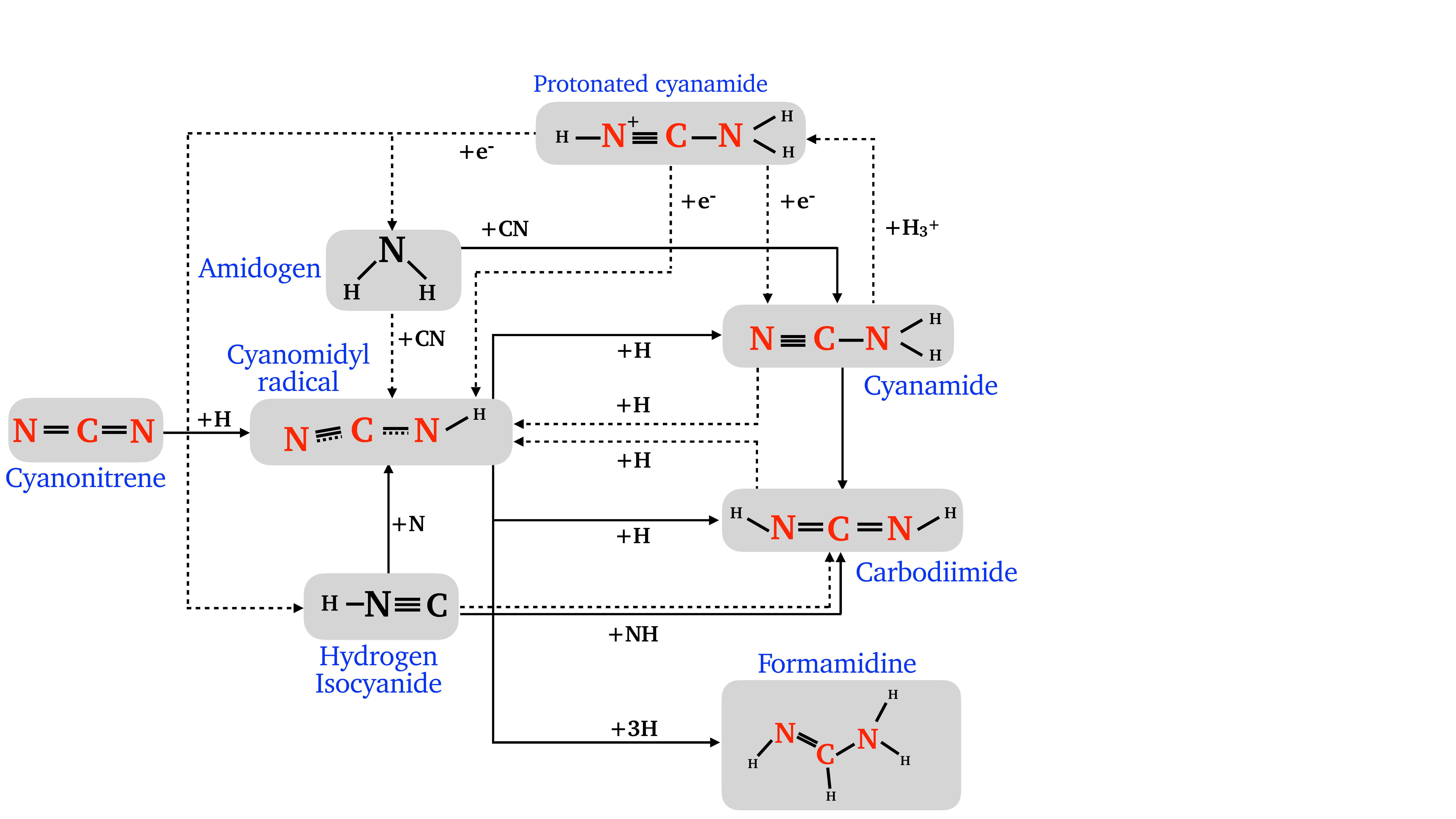}
\vspace{-5mm}
\caption{Chemical network for the formation of species with the NCN backbone (highlighted in red) such as the cyanomidyl radical (HNCN), carbodiimide (HNCNH), cyanamide (NH$_2$CN), protonated cyanamide (NH$_2$CNH$^{+}$) and formamidine (NH$_2$CHNH). Other chemically related species (the NH$_2$ radical and HNC) are also indicated. The name of the different molecules are shown in blue. The solid/dashed arrows indicate the surface/gas chemical routes discussed in the text.} 
\label{fig-diagram}
\end{figure}

Little is known so far about the chemistry of HNCN, mainly due to its lack of detection in the ISM.
We discuss here some formation pathways. The chemistry of the G+0.693 molecular cloud is thought to be mainly dominated by the erosion of the icy grain mantles produced by sputtering in large-scale low-velocity ($\sim$ 20 km s$^{-1}$) shocks present in this region of the Galactic Center (\citealt{zeng2020}). This is reflected in the high abundances of shock tracers like HNCO and SiO (\citealt{martin_tracing_2008,Rivilla2018,zeng2020}), and also of complex species that are known to be formed on icy mantles (e.g. \citealt{requena-torres_organic_2006,zeng2018}). In this context, HNCN could be formed on the grain surfaces, and subsequently injected to the gas phase due to shock-induced desorption. A possible grain-surface pathway would be H-addition to NCN (see Fig. \ref{fig-diagram}), as proposed by the theoretical calculations by \citet{puzzarini2005}, which showed that the process is highly exothermic (89 kcal mol$^{-1}$, 44800 K).
Unfortunately, there are not measurements of NCN abundances on dust grains, and the rotational spectroscopy of NCN is not available.
Another possible grain surface formation route is the N-addition to the carbon atom of hydrogen isocyanide HNC (Fig. \ref{fig-diagram}).

However, HNCN is an unsaturated radical and thus a highly reactive species. As a consequence, one would expect this species to be rapidly converted into more complex species on the surface of dust grains (see Fig. \ref{fig-diagram} and Section \ref{sec:precursor}), making its grain surface abundance basically negligible.
We therefore explore whether this species could be formed by gas-phase chemistry towards G+0.693.

\citet{yamamoto1994} proposed than HNCN can be formed through  electron recombination of protonated cyanamide (NH$_2$CNH$^+$) in the gas phase, yielding NH$_2$CN or HNCN, if one or two atoms are detached during the recombination, or NH$_2$ + HNC, as depicted in Fig. \ref{fig-diagram}. This kind of reactions is in general highly efficient, with typical rate constants of the order of 10$^{-9}$ cm$^3$ s$^{-1}$.
The spectroscopy of NH$_2$CNH$^+$ is not available, so we cannot directly measure its abundance. Other N-bearing protonated species in the ISM are generally less abundant than the neutral ones by 1-2 orders of magnitude (e.g. HCNH$^+$ or HC$_3$NH$^+$, e.g. \citealt{quenard2017}). Thus, the electron recombination of NH$_2$CNH$^{+}$ could be important for the formation of HNCN, which is much less abundant than its putative parent species NH$_2$CN (HNCN/NH$_2$CN$\sim$1/25, Table \ref{tab:parameters}).

HNCN can be regarded as the cyanide derivative of the amidogen radical (NH$_2$). Both NH$_2$ and CN have been detected in Galactic Center molecular clouds with high abundances of $\sim$10$^{-8}$ (e.g. \citealt{vanDishoeck1993,rivilla2019a}), so the reaction of these two radicals might form HNCN.
This reaction is not included in the available chemical databases such as KIDA (Kinetic Database for Astrochemistry, \citealt{wakelam2012}) or UMIST (\citealt{mcelroy2013}). Making an analogy with the reaction OH + CN $\rightarrow$ OCN + H (see KIDA or UMIST), the reaction NH$_2$ + CN could yield HNCN + H (see Fig. \ref{fig-diagram}). 

An alternative route would be the destruction in the gas phase of cyanamide (NH$_2$CN) and/or carbodiimide (HNCNH) by atomic H (Fig. \ref{fig-diagram}), which is highly abundant in the Galactic Center (\citealt{requena-torres_organic_2006}) due to the photodissociation of H$_2$ by secondary UV photons generated by the enhanced cosmic-ray ionization rate in the Galactic Center (\citealt{goto_cosmic_2013}). Unfortunately, these reactions are not included in KIDA or UMIST, and therefore it is not possible to evaluate their efficiency under the physical conditions found toward G+0.693. Future theoretical and experimental works studying these reactions are thus needed.

\subsection{HNCN as precursor of prebiotic species}
\label{sec:precursor}

The first detection of HNCN in the ISM presented in this work led us to consider its possible role as a precursor of prebiotic species. The high reactivity of the HNCN radical suggests that this species could have a key role in the surface formation of other N-bearing species, and in particular in those with the NCN backbone.  

The grain surface hydrogenation onto the N atom bonded with C and H could yield cyanamide (NH$_2$CN), as shown in Fig. \ref{fig-diagram}. The latter species has been proposed as a key precursor of RNA and DNA nucleotides in prebiotic experiments that mimic the conditions on early Earth (e.g. \citealt{powner2009,fahrenbach2017,kaur2019}).
The high NH$_2$CN/HNCN ratio found in G+0.693 around 25 suggests that this hydrogenation could be very efficient, similar to other grain surface hydrogenations such as the one forming H$_2$CO from HCO (\citealt{watanabe2002}). This route, along with the radical-radical surface reaction between NH$_2$ and CN (Fig. \ref{fig-diagram}) proposed by \citet{coutens2018}, might be responsible for the observed high abundance of NH$_2$CN, since there are not known viable routes to form NH$_2$CN in the gas-phase 
(\citealt{talbi2009,blitz2009}).

HNCN could also be hydrogenated on the surface of grains to form the cyanamide isomer carbodiimide (HNCNH, see Fig. \ref{fig-diagram}), as proposed by the calculations by \citet{yadav2019}. This latter species is considered a key condensing agent in the assembling of amino acids into peptides in liquid water (\citealt{hartmann1984}). \citet{yadav2019} also proposed that carbodiimide might be formed (on grain surfaces or in gas-phase) by the reaction between HNC and NH. 
Alternative formation routes of HNCNH have been proposed based on laboratory experiments, such as photo-isomerisation of NH$_2$CN after UV irradiation of ices at low temperatures ($\sim$10 K, \citealt{duvernay2005}), or surface isomerization of NH$_2$CN  in amorphous water ices at T$>$80 K (\citealt{duvernay2004}). 
The latter cannot occur in G+0.693, since the temperatures of the dust grains in Galactic Center clouds are much lower, $T_{\rm dust}\leq$30 K (\citealt{rodriguez-fernandez2004,guzman2015}).
Photo-isomerisation in the ISM can take place in photo-dominated regions (PDR), as proposed to explain the formation of cis-HCOOH from trans-HCOOH in the Orion PDR, where the UV radiation field is about 10$^{4}$ times the interstellar value (PDR, \citealt{cuadrado2016}). In molecular clouds there is a significantly smaller internal UV field of secondary photons induced by the impact of cosmic rays with H$_2$ (\citealt{gredel1989}). In galactic disk molecular clouds with cosmic-ray ionisation rate $\sim$10$^{-17}$ s$^{-1}$, the UV radiation field is about 1000 times less intense than the interstellar UV field (\citealt{caselli2012}), while in Galactic Center molecular clouds (with cosmic-ray ionisation rates of $\sim$10$^{-15}$ s$^{-1}$, \citealt{yusef-zadeh_interacting_2013}), the expected UV field is about 10 times less intense than the interstellar field, and thus five orders of magnitude lower than that of an Orion-like PDR. Hence, photo-isomerisation of NH$_2$CN is an unlikely origin for HNCNH in G+0.693.

We note that HNCNH is at least one order of magnitude less abundant than NH$_2$CN in the gas phase of G+0.693 (Table \ref{tab:parameters}), similarly to what has been found in the Sgr B2N hot core (\citealt{nummelin_three-position_2000,macguire2012}). 
This might indicate that surface hydrogenation of HNCN to NH$_2$CN, which is the most stable isomer, is more efficient than that to HNCNH, and that the isomerization of NH$_2$CN in HNCNH is indeed not efficient in these environments. In any case, gas phase reactions involving both isomers should be studied in the future to understand better their observed relative abundance in the ISM.

Finally, successive surface hydrogenation of the HNCN radical can also lead to formamidine (NH$_2$CHNH, see Fig. \ref{fig-diagram}). This species has been proposed as a main  precursor of amino acids in prebiotic experiments (\citealt{kitadai2018}). Unfortunately, no rotational spectroscopy of this species is available, which prevents its search in the ISM.

\vspace{-6mm}
\section{Conclusions}

We report the discovery in the ISM of the cyanomidyl radical (HNCN) towards the G+0.693-0.027 molecular cloud in the Galactic Center. We have targeted the spectroscopic doublets of the $N$=2$-$1, 4$-$3, 5$-$4, 6$-$5 and 7$-$6 transitions, and detected three unblended lines (namely the $N$=6$-$5 doublet and one $N$=4$-$3 line), and one $N$=5$-$4 line that is only partially blended.
The LTE best fit provides a column density of $N$=(1.2$\pm$0.1)$\times$10$^{13}$ cm$^{-2}$ and an excitation temperature of $T_{\rm ex}$=8.7$\pm$0.7 K. The  abundance with respect to molecular hydrogen is  (0.91$\pm$0.05)$\times$10$^{-10}$.
The unsaturated nature of the HNCN radical makes it highly reactive and thus it likely represents a precursor (or and intermediate step) of other species with the NCN backbone with prebiotic relevance, such as cyanamide (NH$_2$CN), carbodiimide (HNCNH), and formamidine (NH$_2$CHNH), through rapid hydrogenation on the surface of dust grains. This would result in a low abundance of HNCN on ices, and therefore, we propose that the origin of the HNCN detected towards G+0.693-0.027 is likely gas-phase chemistry. Possible gas-phase formation routes would involve electron recombination of protonated cyanamide, the reaction between NH$_2$ and CN, or the destruction of NH$_2$CN and HNCNH by atomic H.

\vspace{2mm}
{\bf Data availability}
The data underlying this article will be shared on reasonable request to the corresponding author.

\vspace{-6mm}
\bibliographystyle{mnras}
\bibliography{hncn}

\vspace{-6mm}
\section{ACKNOWLEDGEMENTS} 

We acknowledge the anonymous reviewer for her/his suggestions that have improved the original manuscript.
V.M.R. thanks Dougal Ritson for interesting discussions about the prebiotic relevance of the species included in this work.
We are grateful to the IRAM 30m and Yebes 40m telescopes staff for their help during the different observing runs.  The 40m radio telescope at Yebes Observatory is operated by the Spanish Geographic Institute (IGN, Ministerio de Transportes, Movilidad y Agenda Urbana).
V.M.R. and L.C. have received funding from the Comunidad de Madrid through the Atracci\'on de Talento Investigador (Doctores con experiencia) Grant (COOL: Cosmic Origins Of Life; 2019-T1/TIC-15379).
I.J.-S. and J.M.-P. have received partial support from the Spanish project numbers PID2019-105552RB-C41 and MDM-2017-0737 (Unidad de Excelencia Mar\'ia de Maeztu$-$Centro de Astrobiolog\'ia, INTA-CSIC).
P.dV. and B.T. thank the support from the European Research Council through Synergy Grant ERC-2013-SyG, G.A. 610256 (NANOCOSMOS) and from the Spanish Ministerio de Ciencia e Innovación (MICIU) through project PID2019-107115GB-C21. B.T. also thanks the Spanish MICIU for funding support from grant PID2019-106235GB-I00.



\clearpage
\appendix

\section{Spectra of CO and isotopologues towards G+0.693-0.027}
\label{app:A}

\begin{figure}
\centering
\includegraphics[scale=0.385]{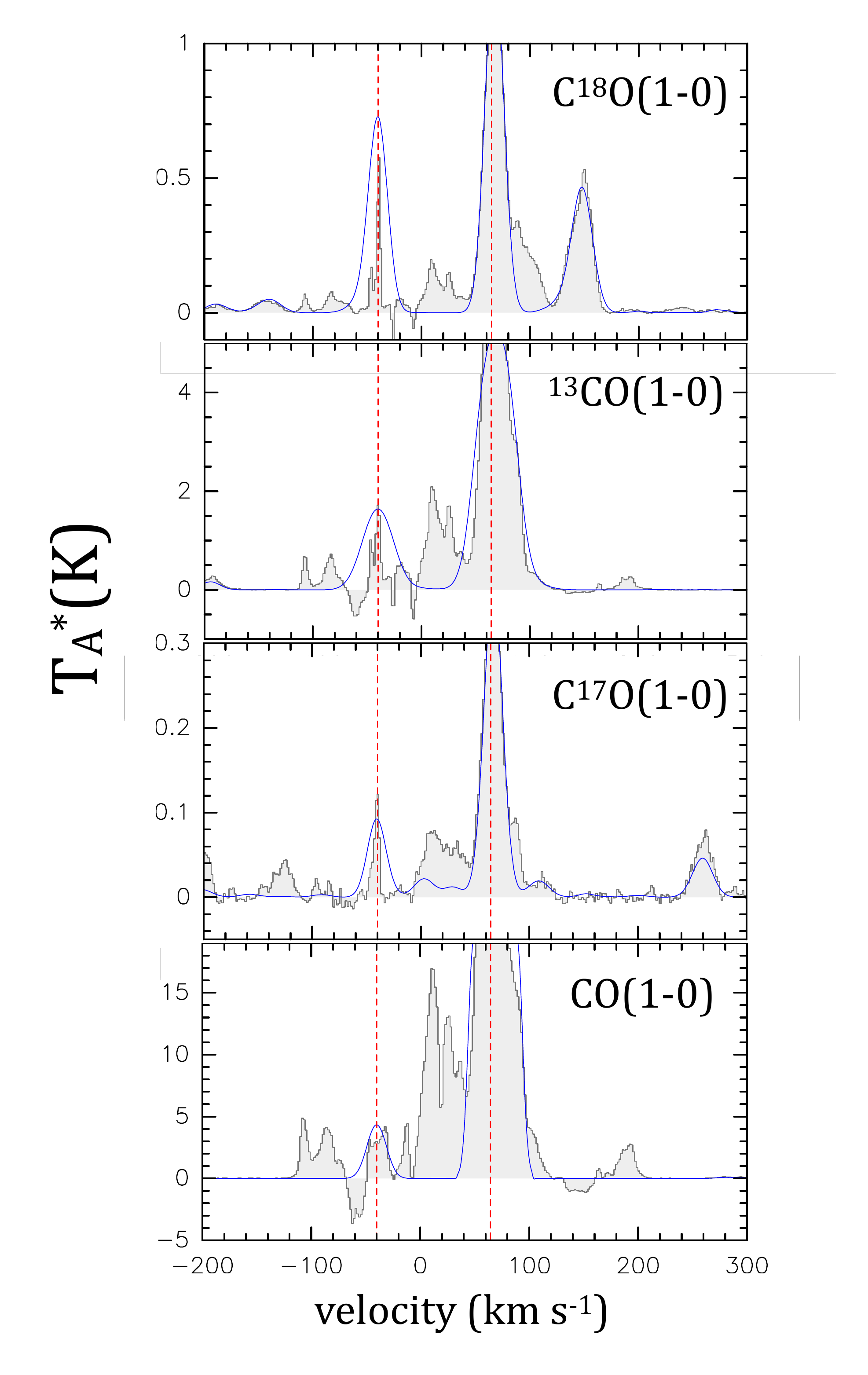}
\centering
\vspace{-5mm}
\caption{Transitions of CO $J$=1$-$0 and its isotopologues towards the G+0.693-0.027 molecular cloud. The blue curves indicate LTE model for the CO and its isotopologues emission (including also other species identified in the cloud). We have considered two velocity components at 69 km s$^{-1}$ and -40 km s$^{-1}$, which are indicated with the vertical dashed red lines. The LTE model for the -40 km s$^{-1}$ component is not a fit but a LTE synthetic spectrum with the same FWHM as the main component at 69 km s$^{-1}$.}
\label{fig-co}
\end{figure}

We show in this Appendix the spectra towards G+0.693 of the $J$=1$-$0 
rotational transition of CO and its isotopologues  $^{13}$CO, C$^{17}$O and C$^{18}$O, observed with the IRAM 30m telescope. The lines show complicated line profiles, with the main emission at the velocity of the G+0.693 cloud at 69 km s$^{-1}$, but also at lower velocities, including a component at $-$40 km s$^{-1}$ (indicated with red vertical dashed lines in Fig. \ref{fig-co}).
The line emission centered at this velocity shows a suspicious narrow velocity component, which is an artefact due to the subtraction of emission from the used off position in the observations, which presents emission from CO and its isotopologues.
The second velocity component of C$^{18}$O at $-$40 km s$^{-1}$ is strongly blended with  the HNCN 5$_{0,5,6}-$4$_{0,4,5}$ transition (see also third row in Fig. \ref{fig-hncn}), which prevents its detection.

\bsp	
\label{lastpage}
\end{document}